\documentclass[traditabstract]{aa}
\usepackage{txfonts}
\usepackage{graphicx}
\usepackage{natbib}

\begin{document}

\title{\textsl{Suzaku} observation of IGR~J16318$-$4848}

\author{\mbox{Laura~Barrag\'an\inst{1}}
\and \mbox{J\"orn~Wilms\inst{1}}
\and \mbox{Katja Pottschmidt\inst{2,3}}
\and \mbox{Michael A.\ Nowak\inst{4}}
\and \mbox{Ingo Kreykenbohm\inst{1}}
\and \mbox{Roland Walter\inst{5}}
\and \mbox{John A.\ Tomsick\inst{6}}
}

\offprints{L.~Barrag\'an,\\ 
(e-mail: laura.barragan@sternwarte.uni-erlangen.de)}

\institute{
Dr.\ Karl Remeis-Sternwarte and Erlangen Centre for Astroparticle
Physics, Friedrich-Alexander-Universit\"at 
Erlangen-N\"urnberg, Sternwartstra\ss{}e~7, 96049 Bamberg, Germany  
\and 
CRESST, University of Maryland Baltimore County, 1000 Hilltop Circle, 
Baltimore, MD 21250, USA 
\and 
NASA Goddard Space Flight Center, Astrophysics Science Division,
Code~661, Greenbelt, MD 20771, USA  
\and 
MIT Kavli Institute for Astrophysics and Space Research, 77,
Massachusetts Avenue, 37-241, Cambridge, MA 02139, USA 
\and 
INTEGRAL Science Data Centre, Geneva Observatory, University of  
Geneva, Chemin d'\'Ecogia 16, 1290
  Versoix, Switzerland 
\and
Space Sciences Laboratory, University of California Berkeley, 7 Gauss
Way, Berkeley, CA 94720-7450, USA
}

\titlerunning{\textsl{Suzaku} Observation of IGR~J16318-4848}
\authorrunning{L. ~Barrag\'an et al.}

\date{Received: --- / Accepted: ---}

\abstract{

  We report on the first \textsl{Suzaku} observation of
  IGR~J16318$-$4848, the most extreme example of a new group of highly
  absorbed X-ray binaries that have recently been discovered by the
  International Gamma-Ray Astrophysics Laboratory (\textsl{INTEGRAL}).
  The \textsl{Suzaku} observation was carried out between 2006 August
  14 and 17, with a net exposure time of 97\,ks.

  The average X-ray spectrum of the source can be well described
  ($\chi_{\mathrm{red}}^{2}=0.99$) with a continuum model
  typical for neutron stars i.e., a strongly absorbed power law
  continuum with a photon index of 0.676(42) and an exponential
  cutoff at 20.5(6)\,keV. The absorbing column is
  $N_\mathrm{H}=1.95(3)\times 10^{24}\,\mathrm{cm}^{-2}$. Consistent with
  earlier work, strong fluorescent emission lines of Fe
  $\mathrm{K}\alpha$, Fe $\mathrm{K}\beta$, and Ni $\mathrm{K}\alpha$
  are observed. Despite the large $N_\mathrm{H}$, no Compton shoulder
  is seen in the lines, arguing for a non-spherical and inhomogeneous
  absorber.

  Seen at an average 5--60\,keV absorbed flux of $3.4\times
    10^{-10}\,\mathrm{erg}\,\mathrm{cm}^{-2}\,\mathrm{s}^{-1}$, the
  source exhibits significant variability on timescales of hours.

}

\maketitle

\section{Introduction}\label{sec:intro}

\object{IGR ~J16318$-$4848} was detected on 2003 Jan.\ 29 during a
scan of the Galactic plane by the IBIS/ISGRI soft gamma-ray detector
onboard the International Gamma Ray Laboratory
\citep[\textsl{INTEGRAL};][]{Courvoisier:03a,Walter:03a}. The source
was the first and most extreme example of a number of highly absorbed
Galactic X-ray binaries discovered with \textsl{INTEGRAL}. Due to the
strong absorption, which can exceed an equivalent hydrogen column of
$10^{24}\,\mathrm{cm}^{-2}$, these sources are extremely faint in the
soft X-rays and had not been detected by earlier missions
\citep{Rodriguez:03b,Patel:04a,Kuulkers:05a}.

Right after its discovery, a re-analysis of archival \textsl{ASCA}
data by \citet{Murakami:03a} revealed a highly photoabsorbed source
($N_\mathrm{H} = 4 \times 10^{23}\,\mathrm{cm}^{-2}$) coincident with
the position given by \textsl{INTEGRAL}. The data also suggested an
iron emission line at 6.4\,keV. These results were confirmed by
various subsequent studies
\citep[e.g.][]{Schartel:03a,Plaa:03a,Revnivtsev:03a,Walter:03a}.
\citet{Matt:03a} detected intense Fe K$\alpha$, Fe K$\beta$, and Ni
$\mathrm{K\alpha}$ emission lines in the spectrum. Based on the
interstellar absorption toward the system, which is two orders of
magnitude lower than the measured $N_\mathrm{H}$,
\citet{Revnivtsev:03b}, \citet{Filliatre:04a}, and
\citet{Lutovinov:05a} also suggested that much of the X-ray absorption
is intrinsic to the compact object.

In an optical study of the system, \citet{Filliatre:04a} proposed that
IGR~J16318$-$4848 is a High Mass X-ray Binary (HMXB) with an sgB[e]
star as the mass donor surrounded by a dense and absorbing
circumstellar material \citep[see also][]{Revnivtsev:03b,moon07a}.
This dense stellar wind results in significant photoabsorption within
the binary system. Based on the optical data, \citet{Filliatre:04a}
suggest a distance between 0.9 and 6.2\,kpc for the system. A likely
location for the source is in the Norma-Cygnus arm
\citep{Revnivtsev:03b,Walter:04a}, which would place it at a distance
of 4.8\,kpc \citep{Filliatre:04a}.

In this \textsl{Paper}, we describe the results of follow-up
observations of IGR~J16318$-$4848 obtained with the \textsl{Suzaku}
satellite, the instruments on which are uniquely suited to study
Compton-thick absorption. In Sect.~\ref{sec:data} we describe the data
reduction. Section~\ref{sec:obs} is devoted to a presentation of the
results of the spectral and temporal analysis. We discuss our results
in Sect.~\ref{sec:conclusions}.

\section{Data analysis}\label{sec:data}

We observed IGR ~J16318$-$4848 with \textsl{Suzaku} from 2006
August~14 until 2006 August~17 for a total net exposure of 97\,ks
(\textsl{Suzaku} sequence number 401094010). We used the standard
procedures to reduce the data from the X-Ray Imaging Spectrometer
\citep[XIS, ][]{Koyama:03a} and the Hard X-Ray Detector \citep[HXD,
][]{Takahashi:07a}. For the XIS in particular we barycentered the
data with \textsl{aebarycen} (version 2008-03-03) and then extracted
source events, images, spectra, and lightcurves with XSELECT v2.4. A
circular source extraction region of 3\farcm23 radius was applied.
The background spectrum was extracted from a circular region having
the same area as the source extraction region. This process was done
for every XIS.  Response matrices and ancillary response files were
generated using XISRMFGEN (version 2009-02-28) and XISSIMARFGEN
(version 2009-02-28), taking into account the hydrocarbon
contamination on the optical blocking filter \citep{Ishisaki:07a}. As
recommended by the \textsl{Suzaku} team, the spectra of the three
front illuminated CCDs (XIS0, XIS2, and XIS3) were then combined with
\textsl{addascaspec} (version 1.30). Although the XIS1 was operational
when the observation was made, it is not used in the present study due
to cross calibration issues.
 
To extract the HXD PIN spectrum, we again followed the standard
procedure of barycentric correction, gti-filtered spectrum extraction
with XSELECT and dead-time correction with HXDDTCOR (version 1.50).
The cosmic background was created with a model provided by the
\textsl{Suzaku} team using a flat response
(ae\_hxd\_pinflate2\_20080129.rsp) and then combined with the internal
background model provided by the \textsl{Suzaku} team
(ae401094010\_hxd\_pinbgd.evt). The resulting combination is used for
the background subtraction. The response matrix used for the analysis
is the one proposed by the \textsl{Suzaku} team for the time of our
observation, ae\_hxd\_pinxinome2\_20080129.rsp. The count rates of
IGR~J16318$-$4848 are
$\mathrm{0.1437\pm0.001}\,\mathrm{cts}\,\mathrm{s}^{-1}$ for the
combined XISs and
$\mathrm{0.6108\pm0.004}\,\mathrm{cts}\,\mathrm{s}^{-1}$ for the HXD
PIN diodes.

For the analysis with XSPEC \citep[v.11.3.2ag;][]{Arnaud:96a} we
rebinned the spectrum to a minimum of 250 and 200\,counts per bin for
the XIS and the PIN, respectively. The uncertainties for all fits are
quoted at the 90\% level for a single parameter of interest. In order
to account for flux cross calibration issues among the instruments, in
all spectral fits a multiplicative constant was introduced. 

\section{Suzaku observation of IGR~J16318$-$4848}\label{sec:obs}
\subsection{Spectral analysis}

\begin{figure}
\resizebox{\hsize}{!}{\includegraphics{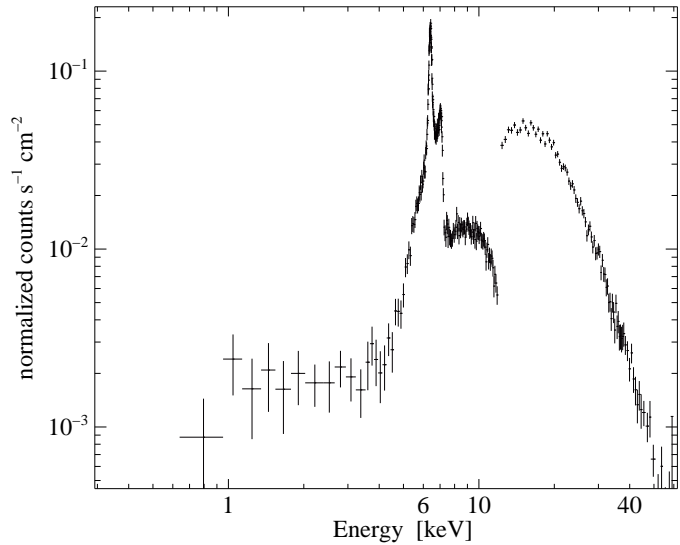}}
\caption{Spectrum of IGR~J16318$-$4848 in the range 0.3--60 keV.}\label{fig:excess}
\end{figure}

\begin{figure}
\resizebox{\hsize}{!}{\includegraphics{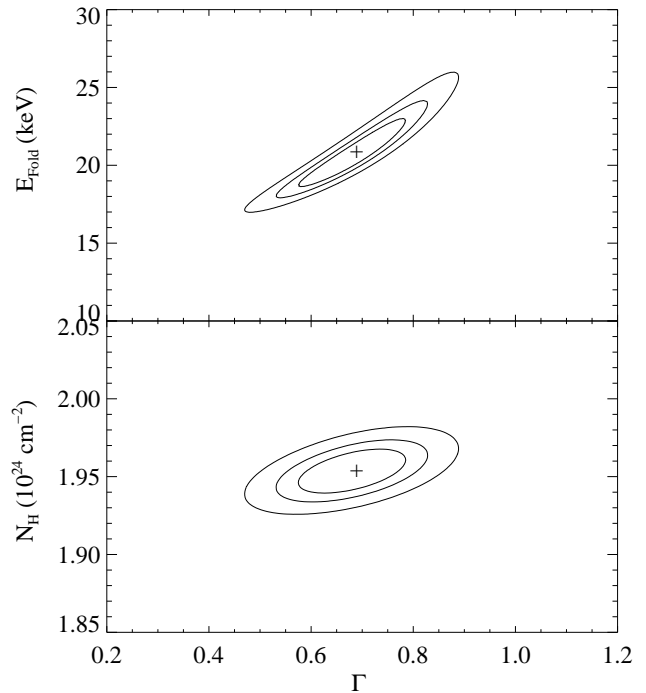}}
\caption{Confidence contours (68, 90, and 99 per cent) of
  the column density and the folding energy as a function of the
  photon index. The cross mark indicates the best fit value.}\label{fig:contour}
\end{figure}

\begin{figure}
\resizebox{\hsize}{!}{\includegraphics{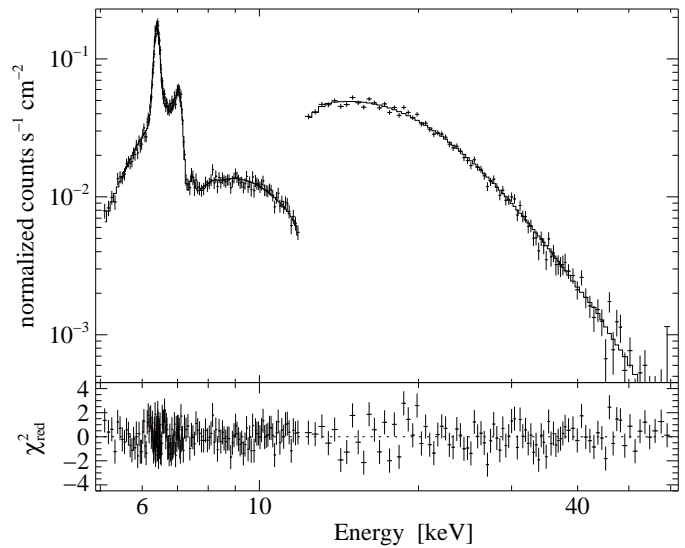}}
\caption{Broad band spectrum of IGR~J16318$-$4848 together with the
  best fit model and its residuals.}\label{fig:spec}
\end{figure}

\begin{figure}
\resizebox{\hsize}{!}{\includegraphics{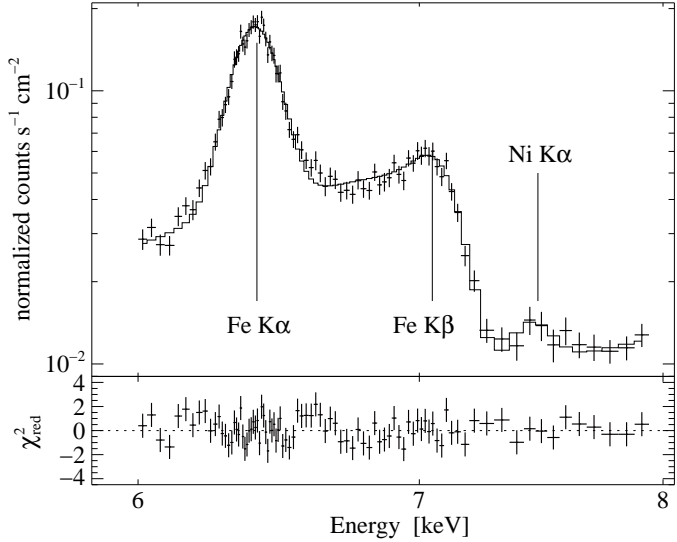}}
\caption{Close-up of the Fe K$\alpha$ band.}\label{fig:closeup}
\end{figure}

\begin{table*}
  \caption{Best fit parameters obtained from modeling the joint XIS
      and HXD data in the 5--60\,keV band.\thanks{$^1$}}
\label{tab:param} 
\begin{tabular}{llll}
\hline\rule{0mm}{3.5mm}
$A_{\mathrm{cutoffpl}}=3.79_{-0.03}^{+0.05}\cdot 10^{-2}$ &
$F_{\mathrm{Fe\ K}\alpha_1}=3.7\pm0.1\cdot 10^{-3}$  &
$F_{\mathrm{Fe\ K}\alpha_2}=1.85\pm0.05\cdot10^{-3}$ \\
&
$F_{\mathrm{Fe\ K}\beta_1}=3.2_{-0.4}^{+0.3}\cdot 10^{-4}$ &
$F_{\mathrm{Fe\ K}\beta_3}=1.57_{-0.20}^{+0.15}\cdot 10^{-4}$ &
$F_{\mathrm{Ni\ K}\alpha}=7.4_{-2.7}^{+2.2}\cdot 10^{-4}$ \\[1mm]
\hline\rule{0mm}{3.5mm}
$c=1.00\pm0.01$ & 
$\Gamma=0.676_{-0.042}^{+0.009}$ &
$E_\mathrm{Fold} =20.5_{-0.3}^{+0.6}$\,keV \\[1mm]
$N_\mathrm{H}=1.95_{-0.03}^{+0.02}\cdot 10^{24}\,\mathrm{cm}^{-2}$ &
$A_\mathrm{Fe} =1.14_{-0.02}^{+0.03}$ \\[1mm]
$E_{\mathrm{Fe\ K}\alpha_1}=6404_{-2}^{+3}$\,eV& 
$\mathrm{EW}_{\mathrm{Fe\ K}\alpha_1}=467_{-54}^{+13}$\,eV&
$E_{\mathrm{Fe\ K}\alpha_2}=6391_{-2}^{+3}$\,eV& 
$\mathrm{EW}_{\mathrm{Fe\ K}\alpha_2}=233_{-27}^{+7}$\,eV\\[1mm]
$E_{\mathrm{Fe\ K}\beta_1}=7093_{-14}^{+13}$\,eV &
$\mathrm{EW}_{\mathrm{Fe\ K}\beta_1}=44.1_{-5.2}^{+1.4}$\,eV &
$E_{\mathrm{Fe\ K}\beta_3}=7092_{-14}^{+13}$\,eV &
$\mathrm{EW}_{\mathrm{Fe\ K}\beta_3}=22.1_{-2.7}^{+0.6}$\,eV \\[1mm]
$E_{\mathrm{Ni\ K}\alpha}=7446_{-51}^{+46}$\,eV& 
$\mathrm{EW}_{\mathrm{Ni\ K}\alpha}=108_{-12.7}^{+4}$\,eV\\[1mm]
\hline\rule{0mm}{3.5mm}
$F_{5.0-60\,\mathrm{keV}}^\mathrm{\,absorbed}=3.4_{-0.1}^{+0.7}\,10^{-10}\,\mathrm{erg}\,\mathrm{cm}^{-2}\,\mathrm{s}^{-1}$ &
$F_{5.0-60\,\mathrm{keV}}^\mathrm{\,unabsorbed}=2.43_{-0.09}^{+0.44}\,10^{-9}\,\mathrm{erg}\,\mathrm{cm}^{-2}\,\mathrm{s}^{-1}$ &
$\chi^2/\mathrm{dof}=242.6/245$ &
$\chi^2_\mathrm{red}=0.99$ \\[1mm]
\hline
\end{tabular}
\end{table*}

Although we detected a soft excess in the spectrum below 5 keV
(Fig.~\ref{fig:excess}), we did not include it in the modeling because
it is most probably due to a serendipitous source at a distance
$\simeq 30''$ from IGR~J16318$-$4848 \citep{Ibarra:07a,Matt:03a}. The
presence of this source could not be confirmed here because of the
lower angular resolution of the XISs compared to \textsl{XMM-Newton},
even when using an optimal attitude solution for \textsl{Suzaku} by
measuring the attitude directly through following the location of
IGR~J16318$-$4848 on the XIS chips.

In order to describe the 5--60\,keV broad-band spectrum of the source
we fit the spectral continuum with an absorbed cutoff powerlaw, taking
also into account non-relativistic Compton scattering. Photoabsorption
was modeled with a revised version of the TBabs model
\citep{Wilms:00a,Wilms:06a}, using the interstellar medium abundances
summarized by \citet{Wilms:00a}. This model describes the continuum
extremely well (Fig.~\ref{fig:spec}). In addition to the continuum,
strong fluorescent emission lines from iron (Fe K$\alpha$ and
K$\beta$) and nickel (Ni K$\alpha$) are introduced in the model
(within the absorber) to obtain a satisfactory description of the data
(Fig.~\ref{fig:closeup}). We model these lines with Gaussians fixed to
a width of $\sigma=0.1$\,eV (i.e., we use lines narrow compared to the
resolution of the XIS). The Fe K$\alpha$ line is modeled as the
superposition of the Fe K$\alpha_1$ and Fe K$\alpha_2$ lines, with the
relative line normalizations held at the 2:1-ratio of the flourescence
yields of these lines and the Fe K$\alpha_2$ line constrained to be
13.2\,eV below the Fe K$\alpha_1$ line. We also modeled the Fe
K$\beta$ line as the combination of the Fe K$\beta_1$ and Fe
K$\beta_3$ lines (the Fe K$\beta_3$ energy being fixed to 16\,eV below
Fe K$\beta_1$, and its intensity to half the one of Fe K$\beta_1$).
This physically correct approach is to be preferred to modeling the Fe
K$\alpha$ and Fe K$\beta$ lines with a single Gaussian. We introduced
a multiplicative constant \textsl{c} to normalize the HXD flux with
respect to the XIS one.

The resulting model (Table~\ref{tab:param}) provides a good
description of the data ($\chi^{2}/\mathrm{dof}=242.6/245$). With
$N_\mathrm{H}=1.95_{-0.03}^{+0.02} \times 10^{24}\,\mathrm{cm}^{-2}$
the column density is very high, as is to be expected for this kind of
source, and is in agreement with the previous observations
\citep[e.g.,][]{Lutovinov:05a,Walter:06a,Ibarra:07a}. In contrast, the
photon index, $\Gamma=0.676_{-0.042}^{+0.009}$, is considerably harder
than in several earlier analyses (e.g., \citealt{Walter:04a}:
$\Gamma=2.6$ or \citet{Ibarra:07a}: $\Gamma=1.35$--1.46). As shown by
the contour plots in Fig.~\ref{fig:contour}, our broad-band data allow
us to determine $\Gamma$ to a high precision. The photon index is not
correlated with $N_\mathrm{H}$, and there is only a slight dependency
between $\Gamma$ and $E_\mathrm{fold}$, which is much smaller than the
difference between the photon index found here and that found in
earlier observations.

Despite the large $N_\mathrm{H}$, which corresponds to a moderately
high Thomson optical depth of $\tau_\mathrm{es}=1.3$, no Compton
shoulder is apparent in the spectrum and all lines are well modeled
with narrow Gaussians (Fig.~\ref{fig:closeup}). In order to determine
an upper limit for the flux in a putative Compton shoulder, following
\citet{Matt:03a} we model this feature by adding a moderately broad
($\sigma=50\,\mathrm{eV}$) Gaussian at 6.3\,keV to the model. The 90\%
upper limit for the flux in the Compton shoulder is
1.8$\times 10^{-5}\, \mathrm{ph}\,\mathrm{cm}^{-2}\,
\mathrm{s}^{-1}$, corresponding to a 90\% upper limit of
34.6\,eV for the equivalent width.

\footnotetext[1]{We list the photon
      index ($\Gamma$), folding energy ($E_\mathrm{Fold}$), hydrogen equivalent
      column ($N_\mathrm{H}$), Fe abundance ($A_\mathrm{Fe}$), the
      total absorbed and unabsorbed fluxes, and the energy ($E$) and
      equivalent width ($\mathrm{EW}$) of the fluorescence lines. The norm
      of the absorbed cutoff powerlaw ($A_{\mathrm{cutoffpl}}$) is
      defined as the photon flux at 1\,keV; for the absorbed Gaussian
      lines the norm ($F$) equals the total line flux.}

\begin{figure}
\resizebox{\hsize}{!}{\includegraphics{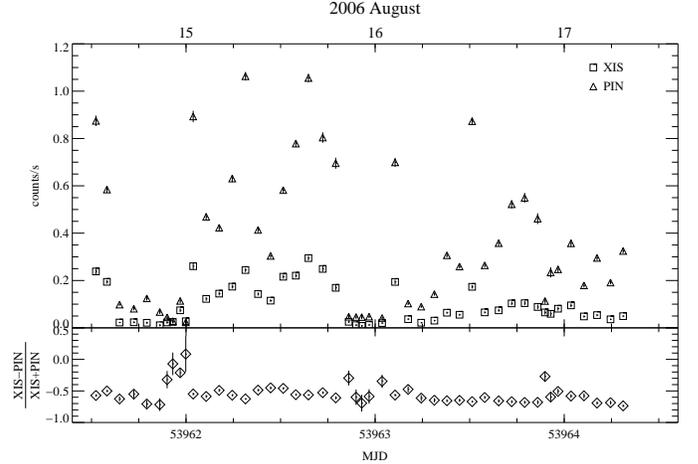}}
\caption{Top: Lightcurve for the XIS (5--12\,keV, squares) and the
  HXD PIN (12--60\,keV, triangles). Bottom: Hardness ratio as a
  function of time.}\label{fig:lightcurve}
\end{figure}

Data from the three XIS and the HXD-PIN were used to obtain
lightcurves in the 5--12\,keV and in the 12--60\,keV band. To
study the evolution of the spectral hardness of the source, count
rates were determined at the resolution of the good time intervals of
the XIS0 detector, which cover approximately one \textsl{Suzaku}-orbit
each ($\sim$90\,minutes). Figure~\ref{fig:lightcurve} shows the
significant variability of IGR~J16318$-$4848 on this resolution.
Throughout the observation, for XIS count rates above
0.1\,$\mathrm{counts}\,\mathrm{s}^{-1}$ the source shows no clear
dependence of the hardness ratio from the source count rate,
indicating that only slight changes in the spectral shape occur. At even
lower count rates, the X-ray spectrum softens, but the signal to
noise in the X-ray spectrum is too low to allow us to quantify these
changes further.

\section{Summary and Conclusions}\label{sec:conclusions}
We have presented first results from the analysis of a long
\textsl{Suzaku} observation of IGR~J16318$-$4848, the most extreme of
the strongly absorbed ``\textsl{INTEGRAL}-sources''. As found in
previous studies, the average spectrum of the source is consistent
with a strongly absorbed exponentially cutoff power-law and strong
flourescent line emission. In contrast to earlier studies, the
power-law photon index was found to be considerably harder than before
($\Delta\Gamma$ from 0.67 up to 1.93). This result can be due to the
significantly better signal to noise ratio in the energy band above
10\,keV compared to the earlier studies, which allows for a better
determination of the high energy cutoff, the continuum parameters, and
$N_\mathrm{H}$ than the earlier soft X-ray measurements, although an
instrinsic change in the source is not ruled out.

The soft excess below 2\,keV is probably due to a serendipitous source
near IGR~J16318$-$4848 \citep{Ibarra:07a}. The considerable
variability of the source can be explained as being due to variations
in $N_\mathrm{H}$.

As pointed out by \citet{Walter:04a}, the general spectral
characteristics derived from the fit are typical for accreting neutron
stars \citep[e.g.,][]{Naik:04a,Hill:08a}. Note that this result does
not mean that the neutron star nature of the compact object in
IGR~J16318$-$4848 is confirmed, which would require e.g. the
detection of pulsations. A search for pulsations in the range between
1\,s and 10\,ksec was negative, while shorter period pulsations are
probably not detectable due to the smearing of pulsations by Compton
scattering \citep{kuster05a}.

Turning to the emission lines, we note that our fit requires a slight
overabundance of iron with respect to the ISM values of
\citet{Wilms:00a}, as one would expect for an evolved star.
Furthermore, the flux ratio of Fe and Ni also points towards a Ni
overabundance by a factor of $\sim$2.5 with respect to Fe.

The ratio of the Fe K$\alpha$ and Fe K$\beta$ line fluxes is given by
$\eta=(F(\mathrm{Fe\ K}\beta_1)+F(\mathrm{Fe\
  K}\beta_3))/(F(\mathrm{Fe\ K}\alpha_1)+F(\mathrm{Fe\ K}\alpha_2))=
0.086\pm 0.008$. This flux ratio is formally slightly smaller than
that found in theoretical calculations for neutral gas phase Fe atoms
of \citet[][$\eta=0.121$]{jacobs86a},
\citet[][$\eta=0.125$]{kaastra93a}, or
\citet[][$\eta=0.132(2)$]{jankowski89a}, and it is also smaller than
the value of $\eta$ found in experimental measurements performed in
solid Fe (e.g., $\eta=0.1307(7)$ found by \citealt{raj98a} and
\citealt{pawlowski02a}). The difference between the different
theoretical calculations is due to certain approximations made in
solving the structure of the excited Fe ion after the K-shell
photoabsorption, while for the latter measurements $\eta$ is affected
by internal absorption in the Fe crystal used to make the measurements
as well as by the dependence of the emission probability of the
photoelectron on orientation. The systematic uncertainty of $\eta$ in
theory and measurements is therefore probably as large as 0.02, which
would make our measurement consistent with neutral Fe. We note that
our value for $\eta$ is significantly smaller than the $\eta
=0.20^{+0.02}_{-0.03}$ found in the \textsl{XMM-Newton} EPIC-pn
analysis of \citet[][but see \citealt{Walter:03a}]{Matt:03a}. These
authors speculated that this higher $\eta$ could be due to the
absorbing wind being moderately ionized. Given that the line ratio
(and also the line energy) found in the higher resolution
\textsl{Suzaku} data are consistent with neutral Fe, we might be
seeing a change in the ionization structure of the wind between the
\textsl{XMM-Newton} and the \textsl{Suzaku} observations.
Alternatively, the larger value for $\eta$ may be due to systematic
effects in the \textsl{XMM-Newton} analysis: With \textsl{Suzaku}, the
Fe K$\beta$ line and the Fe K edge are easier to separate and the
spectral continuum is better constrained in the present analysis than
with \textsl{XMM-Newton}, since spectral information is available
above 9\,keV.

Finally, despite the large column of the source, no significant
evidence for the presence of a Compton shoulder is found in the
\textsl{Suzaku} spectrum, which is consistent with previous results.
This result is in contrast to the expectation for absorption in an
homogeneous medium: As shown by \citet{matt02a}, with this
  assumption the equivalent width of the Fe K$\alpha$ line at the
  $N_\mathrm{H}$ of IGR~J16318$-$484 should be much less than that
observed here, and a strong Compton shoulder should be present, in
line e.g. with the Compton shoulder observed by \citet{Watanabe:03a}
in \object{GX~301$-$2}. As pointed out by e.g.
\citet{Walter:03a,Walter:06a} and \citet{Ibarra:07a}, the
non-existence of the Compton shoulder could be due to a strongly
inhomogeneous absorbing medium. Since the strength of the shoulder is
strongly dependent on the assumed accretion geometry, further work
using self-consistent modeling of the absorption, fluorescent line
formation and Compton shoulder formation is required. We will present
such self-consistent analyses, as well as a more detailed study of the
variability of the source, in a future publication.

\begin{acknowledgements}
  We want acknowledge the anonymous referee for his/her comments that
  allowed us to improve this paper. This work was partially funded by
  the Bun\-des\-mi\-ni\-ste\-rium f\"ur Wirtschaft und Technologie
  through the Deutsches Zentrum f\"ur Luft- und Raumfahrt contract
  50\,OR\,0701 and by National Aeronautics and Space Administration
  grants NNX07AE65G and NNX06AI43G. This research has made use of data
  obtained from the Suzaku satellite, a collaborative mission between
  the space agencies of Japan (JAXA) and the USA (NASA).
\end{acknowledgements}

\end{document}